\documentstyle[aps,preprint,epsfig]{revtex}

\def \beq {\begin{equation}}
\def \eeq {\end{equation}}

\begin{document}

\draft

\title{Chaos around the superposition of a black-hole and a thin disk}

\author{Alberto Saa\footnote{e-mail: asaa@ime.unicamp.br}}
\address{
Departamento de Matem\'atica Aplicada, \\ IMECC--UNICAMP,
        C.P. 6065,\\ 13081-970 Campinas, SP, Brazil}

\author{Roberto Venegeroles\footnote{e-mail: rveneger@fma.if.usp.br}}
\address{
Departamento de F\'\i sica Matem\'atica, \\ Universidade de S\~ao Paulo,
        C.P. 66318, \\ 05315-970 S\~ao Paulo, SP, Brazil}
\maketitle

\begin{abstract}
Motivated by the strong astronomical evidences supporting that huge
black-holes might inhabit the center of many active galaxies, we
have studied the integrability of oblique orbits of test particles
around the exact superposition of a black-hole and a thin disk. We have
considered the relativistic and the Newtonian limits. Exhaustive
numerical analyses were performed, and bounded zones of chaotic
behavior were found for both limits. An intrinsic
relativistic gravitational effect is detected: the chaoticity 
of trajectories that do not cross the disk.
\end{abstract}
\pacs{04.20.Jb, 95.10.Fh, 05.45.+b}

In recent years, strong observational evidences have supported 
that huge black-holes, with masses between $10^6 M_\odot$ and
$10^{10} M_\odot$, might inhabit the center of many active galaxies\cite{kor}.
These evidences have motivated many investigations on 
the black-hole--disk system. Some approximate\cite{beg} and
numerical\cite{lan} results were obtained, and in 
\cite{lele} some exact axisymmetric solutions 
describing  systems containing the superposition of  
non-rotating black-holes
and static thin disks are presented and discussed. 
Such systems have no net angular
momentum, and a possible explanation for their stability is that the
disk particles move under the action of their own gravitational field in such 
a way that there are as many particles moving to one side as to the
other\cite{mm}. This counterrotating interpretation has been
frequently used to describe true rotational effects (see,
for spherical systems, \cite{ein} and, for cylindrical systems,\cite{kip}).

Here, we consider the integrability of oblique orbits of test particles
in the exact static black-hole--disk system. 
There are examples in the literature of chaotic motion
involving black-holes: in the fixed two centers problem\cite{2b},
in a black-hole surrounded by gravitational waves\cite{bw}, 
and in several core--shell models with
relevance to the description of galaxies (see \cite{vl} for a recent
review).
We mention also that geodesic motions
in some static axisymmetric spacetimes were considered in \cite{ssm}, where
chaotic behavior was detected for several solutions containing 
N-point Curzon-like singularities.

We are manly interested in bounded
motions close to the black-hole, so one assumes that the disk is infinite
and has an associated homogeneous Newtonian density.  We have also
considered the Newtonian limit of the black-hole--disk system,
{\em i.e.}, we have also studied the integrability of oblique
orbits of test particles under the Newtonian 
gravitational potential corresponding to the
superposition of a monopole and an infinite homogeneous disk. 
Bounded zones of chaotic
behavior were found for both the relativistic system and for its
Newtonian limit. We also notice that the chaotic regions in the
relativistic cases are typically 
larger than in the corresponding Newtonian ones.
This is a consequence of an intrinsic
relativistic gravitational effect we found: the chaoticity 
of trajectories that do not cross the disk.

Let us start by the Newtonian limit of the black-hole--disk system.
The equations of motion for test particles in this case are very
simple. We use cylindrical coordinates $(r,\theta,z)$ with the monopole,
with mass $M$,
located in the origin. The disk corresponds to the plane $z=0$. The
gravitational potential in this case is given by 
\beq
\label{pot}
V(r,\theta,z) = -\frac{M}{\sqrt{r^2 + z^2}} + \alpha |z|,
\eeq
where $\alpha$ is a positive parameter standing  for 
the mass density of the disk. The
angular momentum $L$ in the $z$ direction is conserved, and we can easily
reduce the three-dimensional original problem to a two-dimensional
one in the coordinates $(r,z)$ and with the Hamiltonian given by
\beq
\label{ham}
H = \frac{\dot{r}^2}{2} + \frac{\dot{z}^2}{2} 
 -\frac{M}{\sqrt{r^2 + z^2}} + \frac{L^2}{2r^2} + \alpha z,
\eeq
for $z>0$. For $z<0$, $H$ is obtained by substituting 
$\alpha\rightarrow -\alpha$.
The Hamiltonian (\ref{ham}) is smooth everywhere except on
the plane $z=0$.
Moreover, the parts of the trajectories restricted to the region
$z>0$ (or $z<0$) are integrable. The Hamilton-Jacobi equations
for a two-dimensional system with the potential
\beq
\label{pot1}
V(r,z) = -\frac{M}{\sqrt{r^2 + z^2}} + \frac{L^2}{2r^2} + \alpha z
\eeq
can be separated in parabolic coordinates\cite{dor}, leading to
the second constant of motion
\beq
C =  R_z - \alpha\frac{r^2}{2} = \dot{r}(z\dot{r} - r\dot{z}) 
-\frac{\alpha}{2} r^2 + \frac{Mz}{\sqrt{r^2+z^2}} + L^2 \frac{z}{r^2},
\eeq
where $R_z$ is the $z$ component of the Runge-Lenz vector
\beq
\vec{R} =  \frac{M}{\sqrt{r^2+z^2}}(r\hat{r}+z\hat{z}) + \vec{v}\times\vec{L},
\eeq
where $\vec{L}$ stands for the total angular momentum.
 In this case, 
with the two constant of motions $H$ and $C$, the equations for the
trajectories of test particles can be reduced to quadratures in
parabolic coordinates\cite{dor}. 
We can take advantage of these results to study the trajectories
of the original system (\ref{ham}). 

Suppose that
a bounded trajectory ($H<0$)
starts somewhere in the region $z>0$. The time evolution of the 
trajectory will be governed by the integrable potential (\ref{pot1}) 
until the plane $z=0$ is reached, say at $r=r_1$, 
$\dot{r}=\dot{r}_1$, and $\dot{z}=\dot{z}_1<0$. After crossing the
disk, the trajectory will be also governed by a 
potential like (\ref{pot1}), but now with a negative parameter $\alpha$.
The trajectory evolves until the plane $z=0$ is reached again, at
$r=r_2$, $\dot{r}=\dot{r}_2$, and $\dot{z}=\dot{z}_2>0$, when 
the signal of $\alpha$ must be changed again in order
to get the correct time evolution, and so on. The
relevant Poincar\'e's section in this case will be given by the 
intersections $\{(r_1,\dot{r}_1),(r_3,\dot{r}_3),\dots\}$.
Since the motion in the regions $z>0$ and $z<0$ are integrable,  
any irregularity in the oblique trajectories shall be credited to the
 signal changes of $\alpha$ between the integrable regions.  
In particular, trajectories that do not cross the disk are integrable,
in contrast to the relativistic case, as we will see latter.

The trajectories in the integrable regions can, in principle, be
described by means of Jacobi elliptic functions. The
expressions are considerably complicated, and we have abandoned
the hope of constructing analytically the Poincar\'e's sections.   
Nevertheless, we could construct them very accurately by solving numerically
the system in the integrable regions and matching the trajectories
appropriately on the disk plane. In the Figures 1 and 2, 
we show typical
Poincar\'e's sections across the plane $z=0$. Fig. 1
presents a low-energy situation $(H=-0.4, L=M=1, \alpha = 0.1)$ where
the integrability seems to be preserved in a large region, while
Fig. 2 presents a section $(H=-0.15, L=M=1, \alpha = 0.1)$
revealing a widespread chaotic behavior. 
We could obtain thousands of intersections for each trajectory
with a cumulative error, measured by the constant $H$,
inferior to $10^{-12}$.

A key point is that the Poincar\'e's sections showed in Fig. 1 and 2
are present for any finite values of $M$ and $\alpha$. In particular,
we will have wide zones of chaotic motion, as presented in Fig. 2,
for any finite values of $M$ and $\alpha$. 
This is a consequence of the invariance of the equations of motion
under the transformations
\begin{eqnarray}
r \rightarrow \lambda r, \quad &M \rightarrow M,& \quad 
z \rightarrow \lambda z \nonumber \\ 
\alpha \rightarrow \lambda^{-2}\alpha, \quad 
&t \rightarrow \lambda^{3/2} t&, \quad 
L \rightarrow \sqrt{\lambda}L,  \\
 & H\rightarrow\lambda^{-1}H   & \nonumber
 \end{eqnarray}
and
\begin{eqnarray}
\label{tr}
r \rightarrow \lambda' r, \quad &M \rightarrow \lambda' M,& \quad
z \rightarrow \lambda' z, \nonumber \\
\alpha \rightarrow {\lambda'}^{-1}\alpha,\quad
&t \rightarrow \lambda' t&, \quad L \rightarrow \lambda' L,\\
 & H\rightarrow H   & \nonumber
 \end{eqnarray}
$\lambda >0$ and $\lambda' >0$. Poincar\'e's sections are invariant, 
up to constant rescalings, under such
kind of transformation, and hence it is possible to choose suitable
values for
$L$ and $H$ in order to have, for instance, the Poincar\'e's section
of Fig. 2 for any finite values of $\alpha$ and $M$.

Now, let us focus the relativistic static black-hole--disk system.
We start with the Weyl metric\cite{kra} describing static axisymmetric
spacetimes,
\beq
\label{weyl}
ds^2 = -e^{\nu(r,z)} dt^2 + e^{\omega(r,z)-\nu(r,z)}\left(
dr^2 + dz^2 \right) + r^2 e^{-\nu(r,z)}d\theta^2.
\eeq
The metric describing the superposition of a black-hole 
with mass $M$ and a
thin disk with homogeneous associated Newtonian density $\alpha$
can be
obtained from the metric of a black-hole immersed in a pseudo-uniform
field, which is described by the potentials
\begin{eqnarray}
\label{pot2}
\nu(r,z) &=& \alpha z + \ln \frac{R_1+R_2-2M}{R_1+R_2+2M}, \nonumber \\
\omega(r,z) &=& -\frac{\alpha^2}{4}r^2 
+\ln \frac{(R_1+R_2-2M)(R_1+R_2+2M)}{4R_1R_2}
+ \alpha(R_2 - R_1),
\end{eqnarray}
where $R_1 = \sqrt{(M-z)^2 + r^2}$ and $R_2 = \sqrt{(M+z)^2 + r^2}$. 
The first terms in the right-handed side of (\ref{pot2}) correspond 
to the pseudo-uniform gravitational potentials, which
can be obtained from the Curzon metric by a limiting process\cite{b}. 
They are called pseudo-uniform
because their Newtonian limit corresponds to a uniform gravitational field.
The potentials (\ref{pot2}) satisfy Einstein vacuum equations
 everywhere except at the origin $r=z=0$. 

One can transform our metric in Weyl coordinates
$(t,r,\theta,z)$ into spherical ones $(t,r_{\rm S},\theta,\varphi)$
by doing
\beq
\label{coo}
r = \sqrt{r_{\rm S}^2 - 2Mr_{\rm S}}\sin\varphi, \quad
z= (r_{\rm S} - M)\cos\varphi.
\eeq
For $\alpha=0$, this coordinate transformation puts the line
element (\ref{weyl}), with the potentials (\ref{pot2}), into
the Schwarzschild form. From (\ref{coo}), one sees directly that
the Schwarzschild radius $r_{\rm S} = 2M$ corresponds in Weyl
coordinates to the rod $r=0$, $-M\le z\le M$,  and that the
circular equatorial photonic orbit at $r_{\rm S}=3M$, $\varphi=\pi/2$,
corresponds to the orbit at $r=\sqrt{3M}$, $z=0$. We note also that
under the transformation (\ref{coo}), any infinitesimally thin disk
in Weyl coordinates remains infinitesimally thin in spherical
coordinates. For more details on the geometry of the Weyl metric,
see \cite{kra} and the references therein.

Our black-hole--disk system corresponds to the juxtaposition of two
black-hole--pseudo-uniform field solutions: the solution (\ref{pot2}) for
$z>0$ and a solution with a reversed uniform field 
$(\alpha\rightarrow -\alpha)$ for $z<0$. The obtained metric 
has $C^0$ components and  
obeys Einstein vacuum equations 
everywhere except in the plane $z=0$. The only
nonvanishing components of the energy-momentum tensor are 
\begin{eqnarray}
\epsilon &=& - T_0^{\ 0} = \alpha 
\frac{e^{\nu(r,z)-\omega(r,z)}}{4\pi}
\left( 1-\frac{M}{\sqrt{M^2 + r^2}}\right)\delta(z), \nonumber \\
p_{\theta\theta} &=& T_3^{\ 3} =  \alpha \frac{e^{\nu(r,z)-
\omega(r,z)}}{4\pi}
\frac{M}{\sqrt{M^2 + r^2}}\delta(z). 
\end{eqnarray}
Our solution obeys the weak energy condition $(\epsilon >0)$
everywhere. The counterrotating velocity $V$ of the particles of the
disk is given by $V^2 = p_{\theta\theta}/\epsilon$\cite{lele}.  
For a relativistically consistent solution, one usually demands that
$V^2\le 1$. For the present case
\begin{equation}
V^2 = \frac{M}{\sqrt{M^2+r^2}-M}.
\end{equation}
We have $V^2\le 1$ for $r\ge \sqrt{3M}$. If we change from the
original Weyl to Schwarzschild
coordinates, we find that it corresponds to $r_{\rm S} \ge 3M$. Hence,
we conclude that our solution is indeed relativistically consistent,
 since it is a well-known 
result that there are no circular orbits inside the photonic 
orbit $(r_{\rm S}=3M$). We are obtaining the consistent
result that the centrifugal-gravity balance inside the photonic
radius can only be maintained for superluminal velocities\cite{lele}.

Now, we need the geodesic equations for the black-hole--disk
system.
Since one always has two independent 
Killing vectors for static axisymmetric spacetimes, 
the geodesic equations for the Weyl metric (\ref{weyl}) can be cast as
\begin{eqnarray}
\label{s1}
\ddot{r} + f_1(r,z) \left(\dot{r}^2 - \dot{z}^2 \right) + 2f_2(r,z)
\dot{r}\dot{z} + g_1(r,z) &=& 0, \nonumber \\
\ddot{z} + f_2(r,z) \left(\dot{z}^2 - \dot{r}^2 \right) + 2f_1(r,z)
\dot{r}\dot{z} + g_2(r,z) &=& 0, 
\end{eqnarray}  
where the dots denote derivation with respect to $s$. For the potentials
(\ref{pot2}), we have:
\begin{eqnarray}
\label{s2}
f_1(r,z) &=& r\left\{ 
\frac{1/R_1+1/R_2}{R_1+R_2+2M} - \frac{1/R_1^2 + 1/R_2^2}{2} -
\frac{\alpha}{2}\left(\frac{1}{R_1} - \frac{1}{R_2}\right) - \frac{\alpha^2}{4}
\right\} \nonumber \\
f_2(r,z) &=& 
{z} \left\{
\frac{1/R_1+1/R_2}{R_1+R_2+2M} - \frac{1/R_1^2 + 1/R_2^2}{2} -
\frac{\alpha}{2}\left(\frac{1}{R_1} - \frac{1}{R_2}\right) 
\right\}  \nonumber \\
&-& M\left\{ 
\frac{1/R_1-1/R_2}{R_1+R_2+2M} - \frac{1/R_1^2 - 1/R_2^2}{2} -
\frac{\alpha}{2}\left(\frac{1}{R_1} + \frac{1}{R_2}\right)
\right\} - \frac{\alpha}{2} \\
g_1(r,z) &=& h_1(r,z)\left[
h_2(r,z)\frac{8M(R_1+R_2)r}{(R_1 + R_2 + 2M)^2}
-\frac{4R_1R_2L^2}{r^3}\exp(2\alpha z)
 \frac{R_1 + R_2 - 2M}{(R_1 + R_2 + 2M)^3}
\right] \nonumber \\
g_2(r,z) &=& h_1(r,z) h_2(r,z) \left[
\frac{8M(R_1(z+M)+R_2(z-M))}{(R_1 + R_2 + 2M)^2} + 2\alpha
R_1R_2\frac{R_1+R_2-2M}{R_1+R_2+2M} \right] \nonumber
\end{eqnarray}
where
\begin{eqnarray}
\label{ss3}
h_1(r,z) &=& \exp\left(\frac{\alpha^2}{4}r^2 + \alpha (R_1-R_2)
\right) \nonumber \\
h_2(r,z) &=& 
\frac{E^2}{(R_1 + R_2 - 2M)^2} +
\frac{L^2}{r^2} \frac{\exp(2\alpha z)}{(R_1 + R_2 + 2M)^2}
\end{eqnarray}
The expressions (\ref{s2}) and (\ref{ss3})  are valid for $z>0$.
For $z<0$, $f_1$, $f_2$, $g_1$, and $g_2$ are
obtained 
by doing $\alpha\rightarrow -\alpha$.
The constants $E=-e^{\nu(r,z)}\dot{t}$ 
and $L=r^2e^{-\nu(r,z)}\dot{\theta}$ are, respectively, the constants of motion
associated to the Killing vectors $\partial/\partial t$ (the energy)
and $\partial/\partial_\theta$ (the $z$-angular momentum). Besides
of these two constants, the requirement of a timelike trajectory
leads to a further one
\begin{equation}
\label{s3}
g_{ab}x^ax^b = 
-1 = e^{\omega(r,z)-\nu(r,z)}\left(\dot{r}^2 + \dot{z}^2 \right)
+\frac{L^2}{r^2}e^{\nu(r,z)} - E^2e^{-\nu(r,z)} 
\end{equation}
We stress here that the system (\ref{s2})-(\ref{s3}) corresponds
to the simplest case of superposition of  black-holes and  disks. 
For the superposition with a non-homogeneous 
or a finite-radius disk, the metric tensor is 
considerably involved\cite{lele}, and even the task of writing down the
geodesic equations for oblique orbits seems to be unrealistic.

We could solve (\ref{s1}) numerically by using the same scheme used
in the Newtonian case. However, we point out a crucial 
difference: the parts of the trajectories contained in the
region $z>0$ (or $z<0$), in contrast to the non-relativistic case,
are themselves non integrable. 
In other words, the trajectories of test particles around
a black-hole immersed in a pseudo-uniform
gravitational field are non integrable. This case 
should be added to the list of known examples of systems presenting
qualitatively distinct behavior in their relativistic and Newtonian
limits\cite{w}, stressing the deep differences between both theories.
This fact clearly indicates that the relativistic
black-hole--disk system is more chaotic than its Newtonian limit, and
we could indeed verify that the chaotic regions in the
relativistic limit are typically 
larger than in the corresponding Newtonian one.

Figure 3 presents a Poincar\'e's section revealing a widespread
chaotic behavior 
for the trajectories around a black-hole immersed in a pseudo-uniform field
($E=0.975$, $L=3.8$, $M=1$, and  $\alpha = 5\times 10^{-4}$).
We could obtain thousands of intersections for each trajectory
with a cumulative error, measured by the constant (\ref{s3}),
inferior to $10^{-12}$. Since the relativistic equations are
invariant under only one type of rescaling, our conclusions
concerning the universality of the Poincar\'e's section presented
in the Fig. 3 are more limited than in the Newtonian case. 
It is not a surprise to find a strong chaotic behavior
of the oblique orbits around the black-hole--disk system, since they are
obtained by matching properly on the disk plane the (chaotic) trajectories
of two black-hole--pseudo-uniform field systems.

We finish discussing our implicit assumption that there is no
interaction between the test particle
and the disk. It was assumed that the test particle may cross the disk
many times without suffering any trajectory deviation. This can be
an acceptable hypothesis if one considers  the disk particles as very
small when compared to the test particle and that we do not have
many crosses. However, even in the case of very small disk particles,
some trajectory deviation is expected to occur after a huge number of 
intersections.
We have considered the following case of a
weak interaction between the disk and the test particle:  
each time the test particle crossed the disk, its velocities
$(\dot{r},\dot{z})$ and $L$ were perturbed by small and  
random amounts such that $H$ (or $E$ in the relativistic case)
was maintained constant. Qualitative 
equivalent Poincar\'e's sections were obtained. However,
a more realistic interaction should also permit small changes
in $H$ (or in $E$). In other words, it would be worth to study the 
behavior of
our models under  more general stochastic 
perturbations\cite{kif}. These points
are now under investigation.

\acknowledgements

The authors are grateful to CNPq, DAAD, and FAPESP for the financial support.
A.S. wishes to thank
Prof. H. Kleinert and Dr. A. Pelster for the warm hospitality at
the Freie Universit\"at Berlin, where this work was
initiated, and Prof. P.S. Letelier for stimulating discussions.

\begin{figure}
\epsfbox{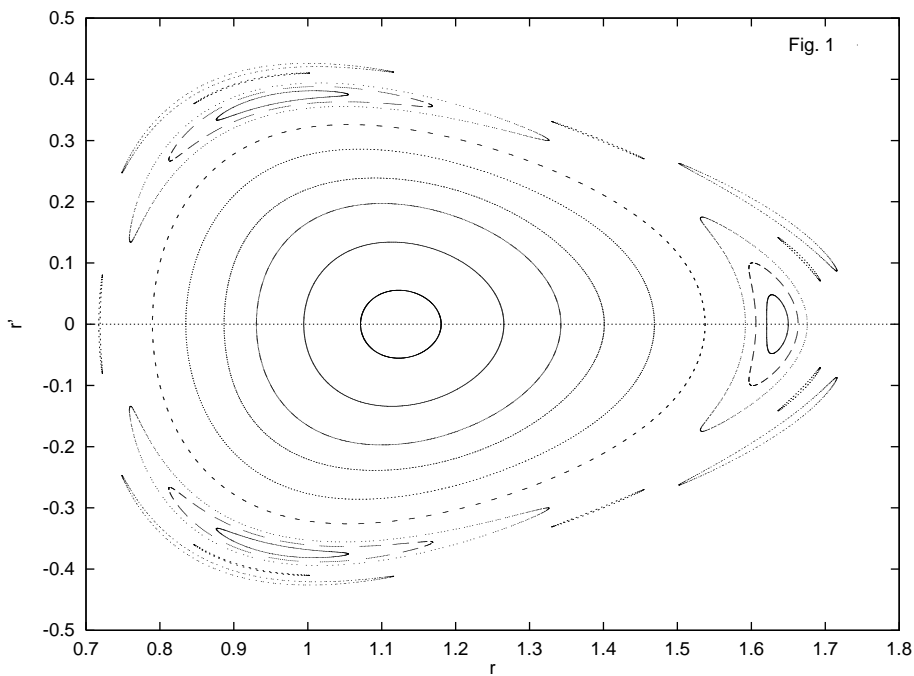}
\vspace{0.5cm}
\caption{Poincar\'e's section $(r,\dot{r})$ across the plane $z=0$ for 
oblique orbits,
with $H=-0.4$ and $L=1$,
 around the superposition of a monopole with
mass $M=1$ and an infinity homogeneous disk with surface
density $\alpha=0.1$.}
\end{figure}

\begin{figure}
\epsfbox{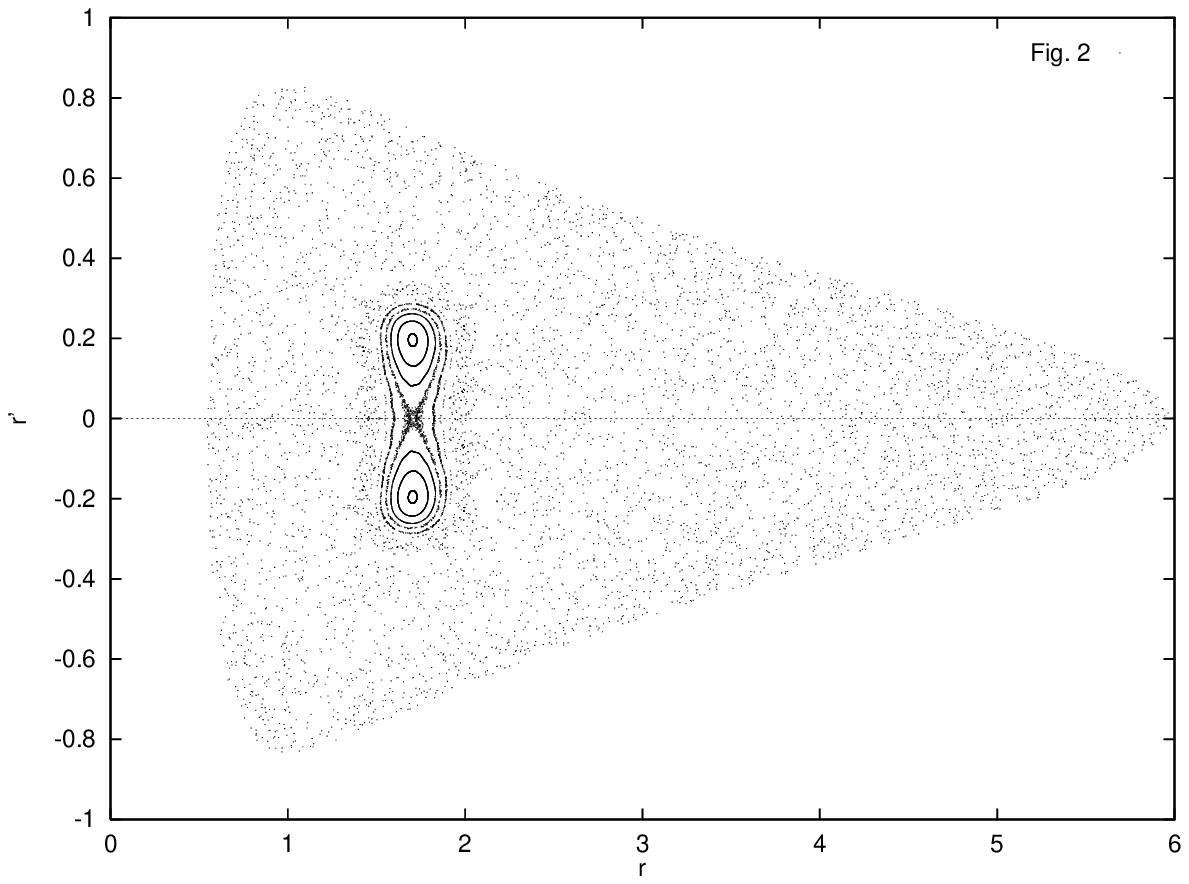}
\vspace{0.5cm}
\caption{Poincar\'e's section $(r,\dot{r})$ across the plane $z=0$ for 
oblique orbits, with $H=-0.15$ and $L=1$, 
around the superposition of a monopole with
mass $M=1$ and an infinity homogeneous disk with surface
density $\alpha=0.1$. }
\end{figure}

\begin{figure}
\epsfbox{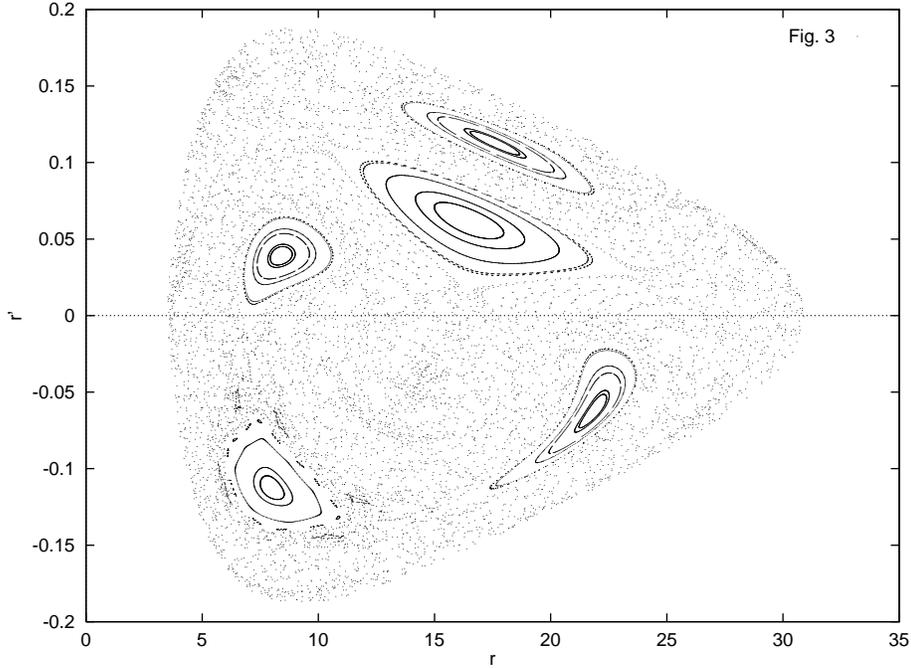}
\vspace{0.5cm}
\caption{Poincar\'e's section $(r,\dot{r})$ across the plane $z=0$ for 
orbits with
$E=0.975$ and $L=3.8$
 around the exact superposition of a black-hole with
mass $M=1$ and a weak pseudo-uniform gravitational field
($\alpha=5\times 10^{-4}$).}
\end{figure}

\end{document}